\documentclass[%
superscriptaddress,
twocolumn,
10pt,
aps,
prb,
floatfix,
]{revtex4-1}

\usepackage[version=4]{mhchem} 
\usepackage{amsmath}
\usepackage{makecell}
\usepackage{amsfonts}
\usepackage{tikz}
\usepgflibrary{shapes.geometric}
\usepackage{color}
\usepackage{subfigure}
\usepackage{graphicx}          %% Include figure files
\usepackage{dcolumn}           %% Align table columns on decimal point
\usepackage{multirow}
\usepackage{bm}                %% bold math
\usepackage{siunitx}
\usepackage{hyperref}
\usepackage{scrextend}
\hypersetup{breaklinks=true,colorlinks=true,citecolor=blue,linkcolor=blue,filecolor=blue,urlcolor=blue}
\usepackage{tabularx}
\usepackage{epsfig}
\usepackage{xcolor,soul}
\usepackage{epstopdf}

\setstcolor{red}    %correction
\bibpunct{[}{]}{,}{n}{}{}   %square brackets for refs

\begin{document}

\title{Theoretical Investigation of Li and Na Oxides Adsorption on TiC(111) Surface for Metal-Air Rechargeable Batteries} 

\author{Keren Raz}
\thanks{Equal contribution author}
\affiliation{School of Chemistry, Tel Aviv University, Israel}
\author{Polina Tereshchuk}
\thanks{Equal contribution author}
\affiliation{Department of Physical Electronics, Tel Aviv University, Israel 69978} 
\author{Diana Golodnitsky}
\affiliation{School of Chemistry, Tel Aviv University, Israel}
\author{Amir Natan}
\email{amirnatan@post.tau.ac.il}
\affiliation{Department of Physical Electronics, Tel Aviv University, Israel 69978}
\affiliation{The Sackler Center for Computational Molecular and Materials Science, Tel-Aviv University, Tel-Aviv 69978, Israel }

\date{\today}

\begin{abstract}
We analyze, with Density Functional Theory (DFT) calculations, the adsorption energies of \ce{Li2O2}, \ce{Na2O2} and \ce{NaO2} on clean and oxygen passivated TiC (111) surfaces. We show, that after deposition of two molecular layers of alkali metal oxides, the initial state of the TiC surface becomes unimportant for the adsorption energy and that all adsorption energies approach their native crystal values. The structure of the adsorbed molecular layers is analyzed and compared to their native oxide crystal structure. Finally, we discuss the possible implications for electrode optimization for Li-air and Na-air batteries.

%Valid PACS numbers may be entered using the \verb+\pacs{#1}+ command.
\end{abstract}


\maketitle

\section{Introduction}
Rechargeable metal-air batteries are widely considered to be the next generation
high-energy-density electrochemical storage devices. The intense interest in
Li-air and Na-air batteries stems from their high theoretical specific energy,
which exceeds that of lithium-ion batteries \cite{Littaucer1997,Abraham1996,Peled2011,
Chakkaravarthy1983,Yang2002,Ma2011,Ito2011,Lu2014,Chen2013,Wen2015,Bruce2012}.
The high specific energy densities of metal-air batteries result from the use of
alkali metals as anodes and ambient-air oxygen, as cathode materials. The
discharge process in Li-air batteries involves an electrochemical reaction between
\ce{Li+} ions and \ce{O2} to form \ce{Li2O2} onto the cathode surface. During charging, oxidation of
lithium peroxide, followed by generation of oxygen gas occurs. 

The optimization of charge and discharge processes involves careful design of the
cathode material, the electrolytes, solvents, and mediators. The selection of solvents 
may affect whether the discharge products are formed at the cathode surface, or in the
solution, and the prevalence of un-wanted side reactions \cite{Bruce2012,Johnson2014}. 

The performance and rechargeability
of metal-air cells strongly depends on the positive electrode material, where oxygen
reduction and evolution reactions take place. 
A suitable cathode material for
an aprotic alkali metal/air cell should have sufficient electronic conductivity,
low density, high stability over the operating voltage of the cathode towards
nucleophilic attack by \ce{LiO2} and O${_2}^{2-}$, low cost, and non-toxicity \cite{Bruce2012}. Carbon
has been the material of choice for the porous cathode. It is known that carbon
oxidizes above 4V versus Li. However, more importantly, carbon raises problems
that impede its use in Li/\ce{O2} cells. Carbon decomposes during oxidation
of \ce{Li2O2} on charging above 3V as a result of the attack by intermediates of \ce{Li2O2} oxidation
and it actively promotes electrolyte decomposition on discharge and charge, rendering
it unsuitable for aprotic Li/\ce{O2} cells \cite{Gallant2012,Ottakam2013}.

Titanium carbide can overcome some of the disadvantages of carbon \cite{Ottakam2013}. Recently, nanocrystalline
TiC has been shown to be an efficient gas diffusion cathode \cite{Ottakam2013}.
TiC has good metallic conductivity. Furthermore, the formation of a passivating
{\it monolayer} of \ce{TiO2} on the TiC surface was reported to be vital to the system's cycleability.
The atomic oxide layer greatly reduces side reactions associated with electrode and electrolyte
degradation at the electrolyte/cathode interface as compared with carbon \cite{Kundu2015,
Ottakam2013,kozmenkova2016tuning}.

While the Li-air battery has the highest theoretical
energy density \cite{Lu2014,Chen2013,Wen2015,Bruce2012,Zu2011}, the low availability of
lithium might lead to future depletion. In contrast to lithium, there are
abundant sodium sources in both the earth's crust (2.3\%) and in the oceans (1.1\%) \cite{Sun2012}.
Moreover, the production of sodium is cheaper than that of lithium. Na-air battery systems have
a lower theoretical specific energy density compared to Li-air battery systems (1605 or 1108 Wh kg-1 considering \ce{Na2O2} or
\ce{NaO2} as discharge products, respectively).
However, Na-air batteries also demonstrate lower charge/discharge overpotential,
which may result in better durability \cite{Peled2011}. Therefore, Na-\ce{O2} battery
offers an interesting alternative to the Li-\ce{O2} battery. Even though sodium and lithium share
many physicochemical properties, the chemistry of the Li-air and Na-air cells is not the same.
While sodium forms stable sodium superoxide, lithium superoxide is thermodynamically unstable
\cite{Sun2012}. It is expected that both sodium peroxide and superoxide would be formed under
different physicochemical conditions, however kinetic factors, temperature and oxygen pressure, the type of support and catalysts
may stabilize a certain phase over the other\cite{Kang2014,Lutz2016}. 

A key process of lithium-air and sodium-air systems is the possible
adsorption of reaction products and intermediates at the cathode surface. The surface adsorption
energy and product growth (e.g., \ce{Na2O2}) at the surface, can be compared to the energetics
of product formation in solution, where both kinetics and \ce{Li+} and \ce{Na+} energetics can affect
the final result. Recently, the adsorption of \ce{Li2O2} clusters on the TiC(111) surface was theoretically
investigated by Wang et al.\cite{Wang2014,Wang2015} using density functional theory. 

In this work we performed ab-initio DFT simulations of $n$\ce{Li2O2}, $n$\ce{Na2O2} and $n$\ce{NaO2}
($n$ = 1$-$6) molecular adsorption on the pristine and oxidized TiC(111) surfaces. We investigated
structural and adsorption energy trends as a function of molecular density on both pristine and oxygen passivated TiC(111) surfaces. The highest magnitude of the adsorption energy was found for
molecules on the pristine surface at $n$ = 1, 2 that bind directly with the reactive Ti-terminated surface,
forming O$-$Li/Na bilayers. At larger molecular density of alkali metal oxides (AMO) there is a reduction in adsorption energy 
and finally after two molecular layers of AMO the adsorption energy gain reaches an almost constant value
close to the AMO native crystal growth energy because of no direct contact with the TiC surface. We also discuss
the work function and d-band shift as a function of AMO surface density. Finally, we discuss the possible implications of our
results for the optimization of TiC and other electrode materials for Li-air and Na-air batteries.

\section{Theoretical Approach and Computational Details}

For the total energy calculations we have used Density Functional Theory (DFT) with the generalized gradient
approximation (GGA) \cite{Perdew1992} functional as proposed by Perdew, Burke, and Ernzerhof (PBE) \cite{Perdew1996}.
Since titanium is a transition metal, one might consider utilization of the Hubbard correction (PBE+$U$). In
addition, Van der Waals (VdW) corrections might affect the adsorption properties of the systems considered.
We analyze both effects in the supporting information (SI) and show that they do not significantly change
our conclusions. We therefore use the PBE functional throughout all the calculations that are shown here.

To solve Kohn-Sham DFT equations we used projected augmented wave (PAW) pseudopotentials \cite{Blochl1994,Kresse1999} as
implemented in the Vienna Ab-initio Simulations Package (VASP) \cite{Kresse1996,Hafner2008}. We used a
Monkhorst-Pack \textbf{k}-point sampling scheme \cite{Monkhorst1976} with \textbf{k}-point meshes of $11\times11\times11$,
$11\times11\times1$ and gamma point for the bulk, surface and gas-phase molecules, respectively, and a cutoff
energy of 500~eV for all the calculations, except for the bulk calculations, in which a higher cutoff was used for the
stress tensor minimization procedure. The structures were considered as optimized to their ground state
geometries when the atomic forces per atom were smaller than 0.02~eV~per~{\AA} and a total energy convergence of
$10^{-6}$~eV was achieved.

To model the TiC, \ce{Li2O2} and \ce{Na2O2} bulks and surfaces we used the AFLOW \cite{Curtarolo2012} package
and the ICSD database \cite{ICSD}.
The TiC(111) surface was constructed by applying the repeated slab model with 8 layers, $2\times2$ surface
unit cell and 21~{\AA} vacuum region, which was found to be sufficient. Thicker slabs of TiC were shown to give similar results and are discussed in the SI. We found that the Ti-terminated TiC(111) surface 
is more stable than the C-terminated TiC(111) surface, which is supported by previous results
\cite{Zaima1985,Vojvodic2006,Wang2014}. Thus, for our further simulations of
the molecules on TiC(111) we use the Ti-terminated TiC(111) surface. The oxidized TiC(111) surface was modelled
by covering the Ti-terminated TiC(111) surface by oxygen atoms at the positions of the next imaginary layer of carbon
atoms layer. Our calculated surface energy for the Ti-terminated TiC(111) surface, 201.8~meV/{\AA$^2$},
is slightly lower than the result reported by {\it Wang et. al.} (208~meV/\AA$^2$) \cite{Wang2014}.

\subsection{Atomic structure generation of $n$\ce{M2O2} (M = Li, Na) and $n$\ce{MO2} (M = Na) ($n$ = 1--6) 
on TiC(111) surface}

The adsorption geometry of the AMOs is first dictated by the reactive TiC surface. As more AMO layers are formed, the geometry becomes less affected by the TiC surface and should eventually approach that of the native AMO crystal. Hence, the resulting geometry of the first layers is somewhere in between the positions dictated by the TiC surface and the AMO native crystal. This gradual change makes the finding of a global minimum for the geometry a greater challenge. We 
simulated, as an initial guess, different $n$\ce{Li2O2}, $n$\ce{Na2O2}, and $n$\ce{NaO2} ($n$ = 1--6) structures 
on the pristine and oxidized TiC(111) surfaces by applying the following procedures which are described below
in the text: (i) first-principles molecular dynamics (MD) simulations, (ii) finding a common crystalline cell
and (iii) cross-check procedure. In all of the calculations we placed the molecules on one side of the slab and
employed dipole correction in order to obtain accurate total energies. We allowed the molecules to relax along 
with four slab layers, while the remaining bottom layers were frozen.

The details of the simulations are as follows:

(i) {\bf MD simulations:} we performed MD simulations by applying the Nos\'{e} thermostat and slowly lowering the
temperature from 300 K (and 500 K) to 0 K for 30 ps. The initial structural models were built by putting the optimized
gas-phase \ce{Li2O2}, \ce{Na2O2} and \ce{NaO2} molecules at random positions at about $3-4$~{\AA}
above the pristine and oxidized TiC(111) surfaces.

(ii) {\bf Common crystalline cell:} In order to investigate the crystal \ce{Li2O2}, \ce{Na2O2}, and \ce{NaO2} structures on the
TiC(111) surfaces we employed the geometrical lattice match algorithm proposed
by {\it Zur and McGil} \cite{Zur1984}. This algorithm can be successfully used to fit superlattices with any pair of 
crystals and surfaces. Thus, we created a common cell for the hexagonal \ce{Li2O2}(001) and
\ce{Na2O2}(001) surfaces on the hexagonal TiC(111) surface, and cubic \ce{NaO2}(001) surface on the hexagonal
TiC(111) surface.

It can be shown that 2\ce{Li2O2}(001) and 2\ce{NaO2}(001) form a full layer at the ($2\times2$) TiC(111) surface hexagonal unit cell,
and so we built the crystal structures with $n$ = 2, 4 and 6, which represent one, two and three
crystal layers, respectively. For the case of \ce{Na2O2}(001) crystal, a full monolayer coverage occurs at $n$ = 1.5,
which implies the coverage of 2 and 4 crystal layers at $n$ = 3 and 6, respectively.
This happens because although the ($2\times2$) TiC(111) surface unit cell is commensurate with the \ce{Na2O2} crystal,
it cannot accommodate an integer number of molecules for a single and full molecular layer of its native crystal.
In order to achieve complete crystal layer coverage by an integer number of molecules, we repeated our calculations
with a larger rectangular surface cell that has twice the surface area. The results of these calculations
can be found in SI and generally agree with the analysis of the smaller cell.

(iii) {\bf Cross-check procedure:} Finally, we performed cross-check between the lowest energy structures of $n$\ce{Li2O2}
and $n$\ce{Na2O2} with the aim of investigating additional possible configurations. We took the lowest energy $n$\ce{Li2O2} 
structure and used it as a starting point for geometrical relaxation of the respective $n$\ce{Na2O2} system, and vice-versa. This was helpful in some particular
cases to find lower energy structures that were missed by the MD procedure.

Thus, we constructed about 4-5 initial configurations for every system, which were finally optimized with the 
standard optimization procedure applying the conjugated gradient algorithm as implemented in VASP.
Finally, we collected all these structures and selected the lowest energy configurations, which were
used for further analyses.

\section{Results}
\subsection{TiC, \ce{Li2O2}, \ce{Na2O2} and \ce{NaO2} bulks, common cells and gas-phase molecules}
Titanium carbide (TiC) bulk has a face centered cubic (fcc) structure with the $O_{h}^{5}$ space group 
symmetry. Our calculated TiC equilibrium lattice constant, $a=4.337$~{\AA}, is close to the thermal 
expansion experimental value (4.318~{\AA}) \cite{Elliott1958} and recent PBE calculations (4.333{~\AA}) 
\cite{Wang2015}. \ce{Li2O2} and \ce{Na2O2} bulks crystallize in hexagonal $P63/mmc$ and $P62m$ space
groups, respectively. The lattice parameters we obtained are 3.158~{\AA} and 7.686~{\AA} for \ce{Li2O2}
and 6.195~{\AA} and 4.472~{\AA} for \ce{Na2O2}, in good agreement with the experimental values \cite{Cota2005}
and previous theoretical GGA results \cite{Radin2012,Wang2015,Araujo2015}. \ce{NaO2} can 
crystallize in a pyrite structure in $Pa3$ space group  (between 196 and 223 K), and in $Fm3m$ space group
(above 223 K), which corresponds to a pyrite structure with a disorder of \ce{O2} orientation, as obtained by
powder and single crystal X-ray diffraction methods \cite{Carter1953}. Our calculated structure corresponds 
to the ordered pyrite structure in $Pa3$ space group with a calculated lattice parameter of 5.521~{\AA}, 
which is in a good agreement with other PBE calculations (5.509~{\AA}) \cite{Arcelus2015} and
experiment \cite{Carter1953}. The lattice parameters calculated by us along with the corresponding literature
values can be found in the SI.

For the common cells created, \ce{Li2O2}(001)/TiC(111), \ce{Na2O2}(001)/TiC(111) and \ce{NaO2}(001)/TiC(111), 
we calculated a misfit factor $\eta = (1-\frac{2\Omega}{\Omega+A})*100\%$, as defined by {\it Wang et al.} 
\cite{Wang2015}, and an area ratio $\alpha = \frac{A}{\Omega}$, where $\Omega$ is the surface 
area of TiC(111) (32.597~{\AA}$^2$) and $A$ is the original surface area of \ce{Li2O2}(001) (34.543~{\AA}$^2$),
\ce{Na2O2}(001) (33.237~{\AA}$^2$) and \ce{NaO2}(001) (30.484~{\AA}$^2$). We found a compression and a
relatively small $\eta$ of 2.90\% and 0.97\% and $\alpha$ of 1.06 and 1.02 for the \ce{Li2O2}(001)/TiC(111) 
and \ce{Na2O2}(001)/TiC(111), respectively, and an expansion with $\eta$ of $-3.35$\% and $\alpha$ of 0.94
for the \ce{NaO2}(001)/TiC(111) cell.

To obtain the lowest energy structures of the molecules in the gas-phase, we selected reasonable geometries, such as
linear, square, triangle and rhombus for \ce{Li2O2} and \ce{Na2O2} molecules, and
linear and triangle for the \ce{NaO2} molecule to be optimized. We found that the lowest energy structures for the \ce{Li2O2}
and \ce{Na2O2} molecules are the planar rhombus, in which two O and two Li (Na) atoms are at opposite corners
with Li$-$O (Na$-$O) bond lengths of 1.74~{\AA} (2.08~{\AA}) and O$-$O bond lengths of 1.59~{\AA} (1.60~{\AA}).
This structure is in a good agreement with the results reported by {\it Lau et al.} \cite{Lau2012} for the \ce{Li2O2} molecule. The gas-phase
structure of \ce{NaO2} corresponds to the isosceles triangle with Na$-$O and O$-$O bond lengths of 2.13~{\AA}
and 1.37~{\AA}, respectively. A figure that shows the lowest energy geometry for the gas-phase molecules is presented in the SI.

\subsection{Adsorption energies}
We have calculated the adsorption energy per molecule as:

\begin{equation}
	\label{eq:adsorption_energy_1}
	E_{ad1}^n = (E_{tot}^{\text{n*mol/TiC}} - nE_{tot}^{\text{molecule}} - E_{tot}^{\text{TiC}})/{n}~
\end{equation}
where $n$ is the number of molecules in the system and $E_{tot}^{\text{n*mol/TiC}}$, $E_{tot}^{\text{molecule}}$ and $E_{tot}^{\text{TiC}}$
correspond to the total energies of the lowest energy structures for: $n$\ce{M2O2} and $n$\ce{MO2} molecules on
the non-oxidized and oxidized TiC(111) slabs, a single gas-phase molecule, and the TiC slabs,
respectively.

In order to evaluate the contribution of the recently adsorbed molecule, we calculated also:

\begin{align}
E_{ad2}^n  &=E_{tot}^{\text{n*mol/TiC}}-E_{tot}^{\text{(n-1)*mol/TiC}}-E_{tot}^{\text{molecule}} 
\notag\\ & = nE_{ad1}^{n} - (n-1)E_{ad1}^{n-1}.
\label{eq:adsorption_energy_2}
\end{align}

Here $E_{ad1}^{n}$ and $E_{ad1}^{n-1}$ are the adsorption energies of the $n$ and $n-1$ molecules on the TiC(111) surfaces 
as defined in Eq. \ref{eq:adsorption_energy_1} and we define $E_{ad1}^0=0$.

\begin{figure}
\includegraphics[width=0.45\textwidth]{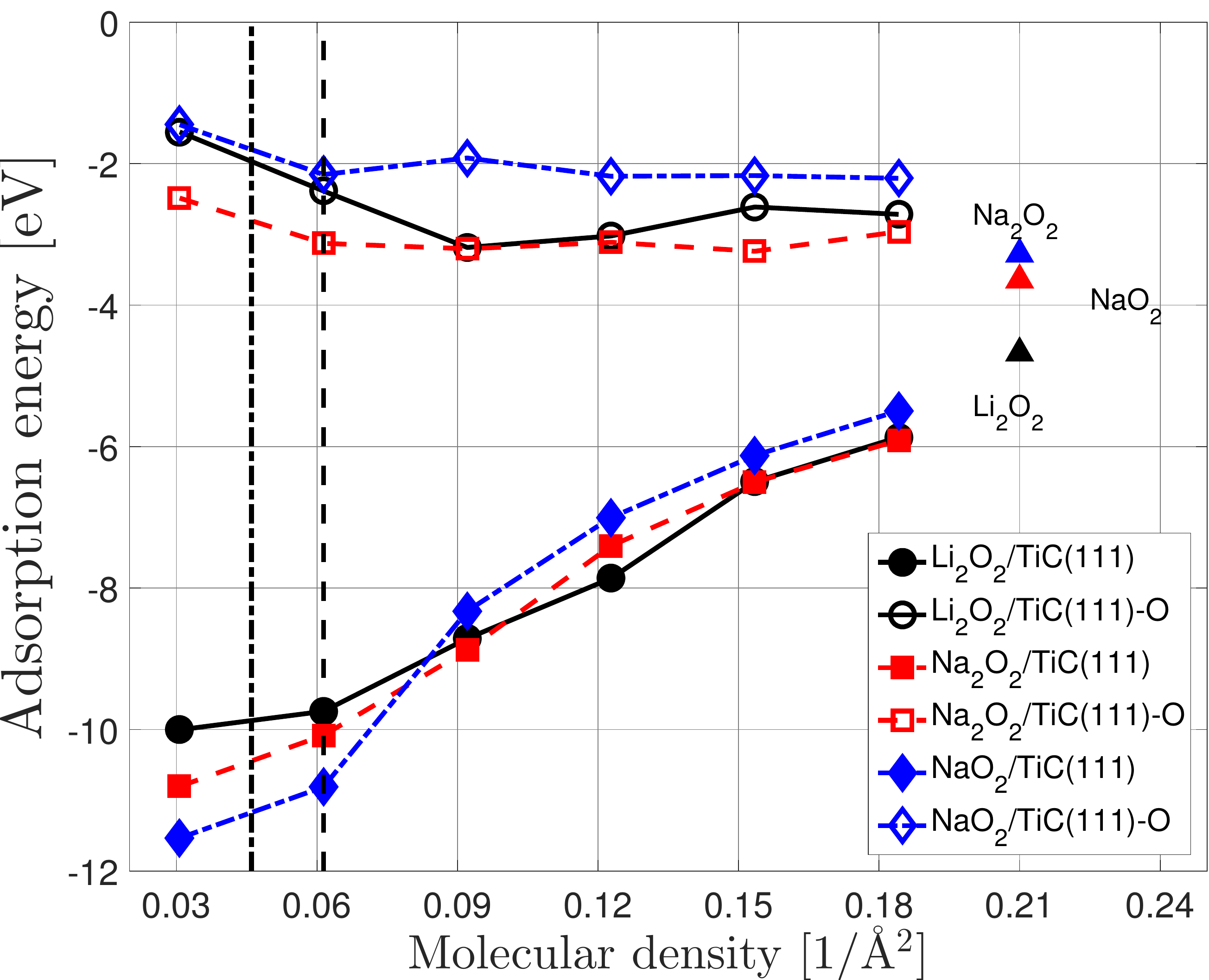}
\caption{\label{fig:energy} The adsorption energy (Eq. \ref{eq:adsorption_energy_1}) of $n$\ce{Li2O2},
$n$\ce{Na2O2} and $n$\ce{NaO2} molecules on TiC(111) and TiC(111)-O surfaces
vs molecular density. Triangles correspond to the growth energy per molecule of the metal oxide crystals 
that were adapted to fit the TiC(111)
surface cell. The dashed vertical lines correspond to the molecular densities that lead to a single molecular layer, i.e., 2 molecules of \ce{Li2O2} and \ce{NaO2} and
1.5 molecules of \ce{Na2O2} in the TiC surface cell.
}
\end{figure}

We calculated the adsorption energies of systems with $n=1$ to $n=6$ molecules corresponding to molecular surface 
densities of $\sim 0.03$[1/\AA$^2$] ($3\times 10^{14}$[1/cm$^2$]) to $\sim 0.18$[1/\AA$^2$] ($1.8\times 10^{15}$[1/cm$^2$]). 
Those results are shown in Figure \ref{fig:energy}, we found similar trends in the adsorption energies ($E_{ad1}$) of $n$\ce{Li2O2}, $n$\ce{Na2O2} and
$n$\ce{NaO2} molecules on the TiC(111) and TiC(111)-O surfaces. For example, $E_{ad1}$ decreases with
increasing coverage of molecules on the TiC(111) surface, from $-10.00$~eV to $-5.87$~eV
for \ce{1Li2O2} and \ce{6Li2O2}, respectively. This trend can be explained by the following argument: 
The larger value of $E_{ad1}$ at $n$=1,2 is due to the direct binding of the molecules
with the highly reactive Ti surface atoms. When increasing the number of molecules in the system,
$E_{ad1}^n$ tends to decrease as the additional molecules are adsorbed on top of the first molecular 
layer and are not in the direct contact with the TiC surface. Since the interaction between molecular 
layers is weaker than the interaction of the molecules with the clean and reactive TiC surface, 
a smaller adsorption energy is obtained.

\begin{figure}
\includegraphics[width=0.45\textwidth]{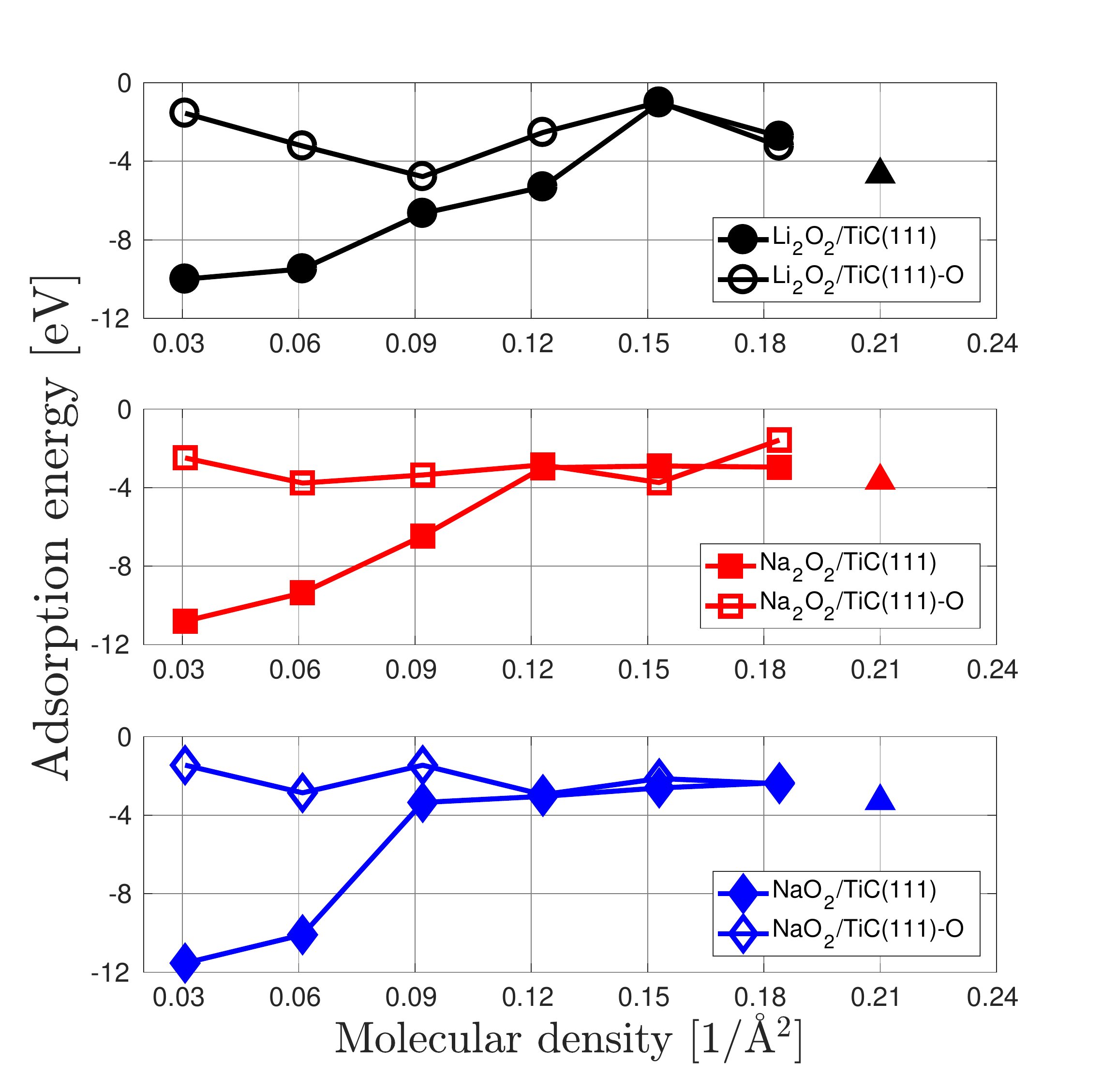}
\caption{\label{fig:adsorption_energy2} The adsorption energy (Eq. \ref{eq:adsorption_energy_2}) of \ce{Li2O2}, \ce{Na2O2} and
\ce{NaO2} on TiC(111) and TiC(111)-O surfaces per molecular density. Triangles
correspond to the growth energy per molecule of the metal oxide crystals in the TiC(111) surface cell.
}
\end{figure}

By contrast, on the oxidized surface, the $E_{ad1}$ of molecules follow a different trend and are in the
range of $-1.55$~eV to $-3.19$~eV for $n$\ce{Li2O2}, from $-2.48$~eV to $-3.24$~eV for $n$\ce{Na2O2}
and $-1.45$~eV to $-2.21$~eV for $n$\ce{NaO2} systems. This behavior shows that the oxygen layer effectively passivates the TiC surface. The AMO molecules now interact with the oxygen layer and not with the reactive Ti atoms of the pristine surface, leading to significantly lower adsorption energies for the first AMO layer.  
Equation \ref{eq:adsorption_energy_2} gives the adsorption energy of the recently added molecule; these data are shown in Figure \ref{fig:adsorption_energy2}.
The adsorption of one and two molecules to the pristine TiC(111)
surface yields more energy because of direct binding with the reactive TiC(111) surface, i.e. $E_{ad2}$ = $-10.00$~eV
for \ce{1Li2O2}, while the adsorption of the next molecules require less energy, e.g. $-6.66$~eV for \ce{3Li2O2},
as the molecules are above the first bilayer. Finally, at $n$ = 5 and 6 the adsorption energy gain remains
unchanged, namely, $-2.73$~eV for \ce{6Li2O2}, and tends to achieve the adsorption energy of the crystal structure,
which is to be expected as a result of the increasing role of the intermolecular binding. Another important observation 
from Figure \ref{fig:adsorption_energy2} is that after two molecular layers (density of 0.12 [1/\AA$^2$]), the adsorption energy for additional molecules is almost the same for the clean and oxidized surfaces for all the three AMO species.  

\subsection{Geometry of adsorbed AMO layers}
\begin{figure*}
\includegraphics[width=0.9\textwidth]{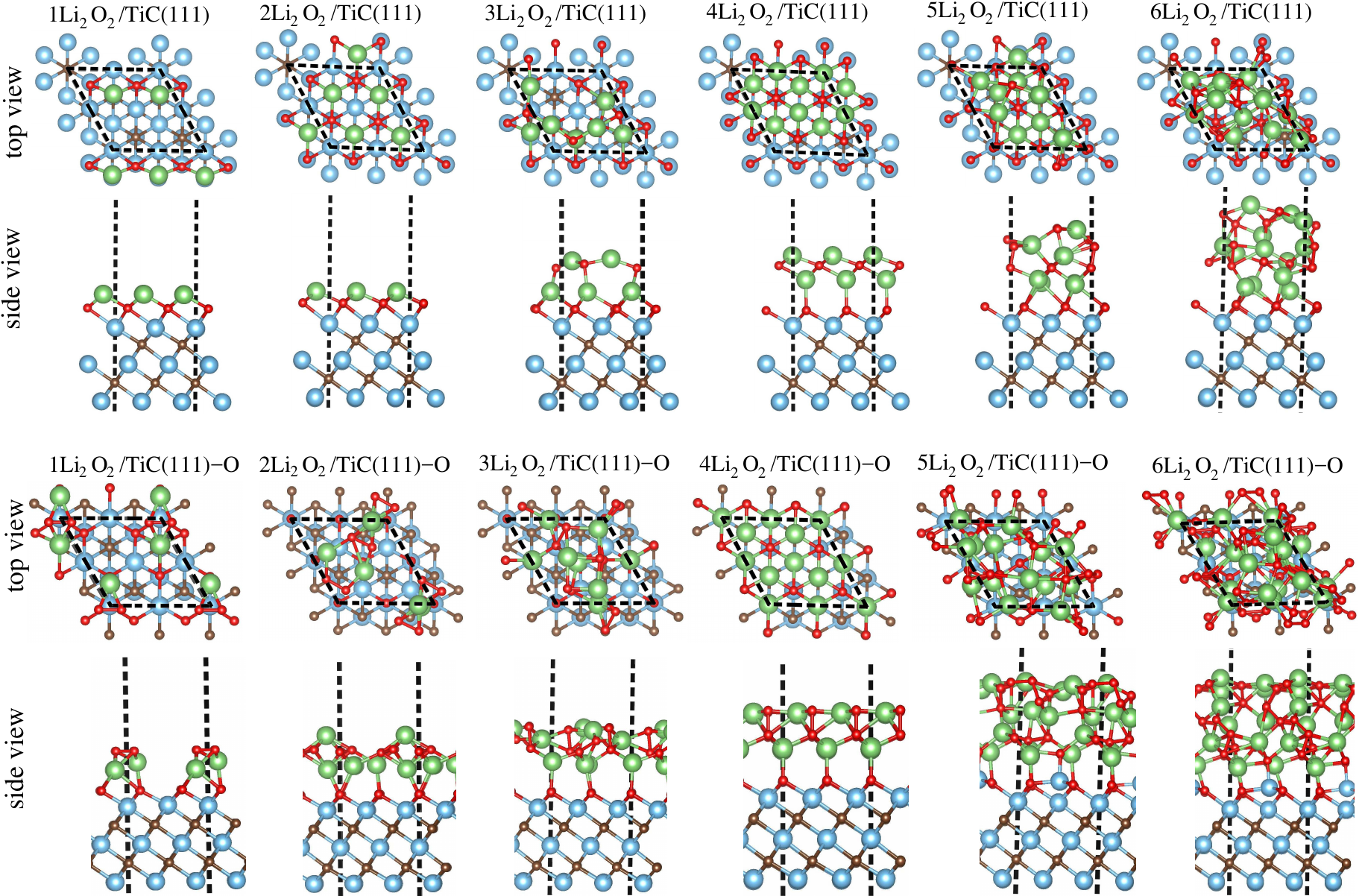}
\caption{\label{fig:Li2O2}Lowest energy configurations of $n$\ce{Li2O2} ($n$ = 1--4) molecules on
the non-oxidized and oxidized TiC(111). Blue and brown balls correspond to Ti and C atoms, while red and green balls are O and Li atoms, respectively. Ball sizes are drawn according to atomic radii.
}
\end{figure*}

\begin{figure*}
\includegraphics[width=0.9\textwidth]{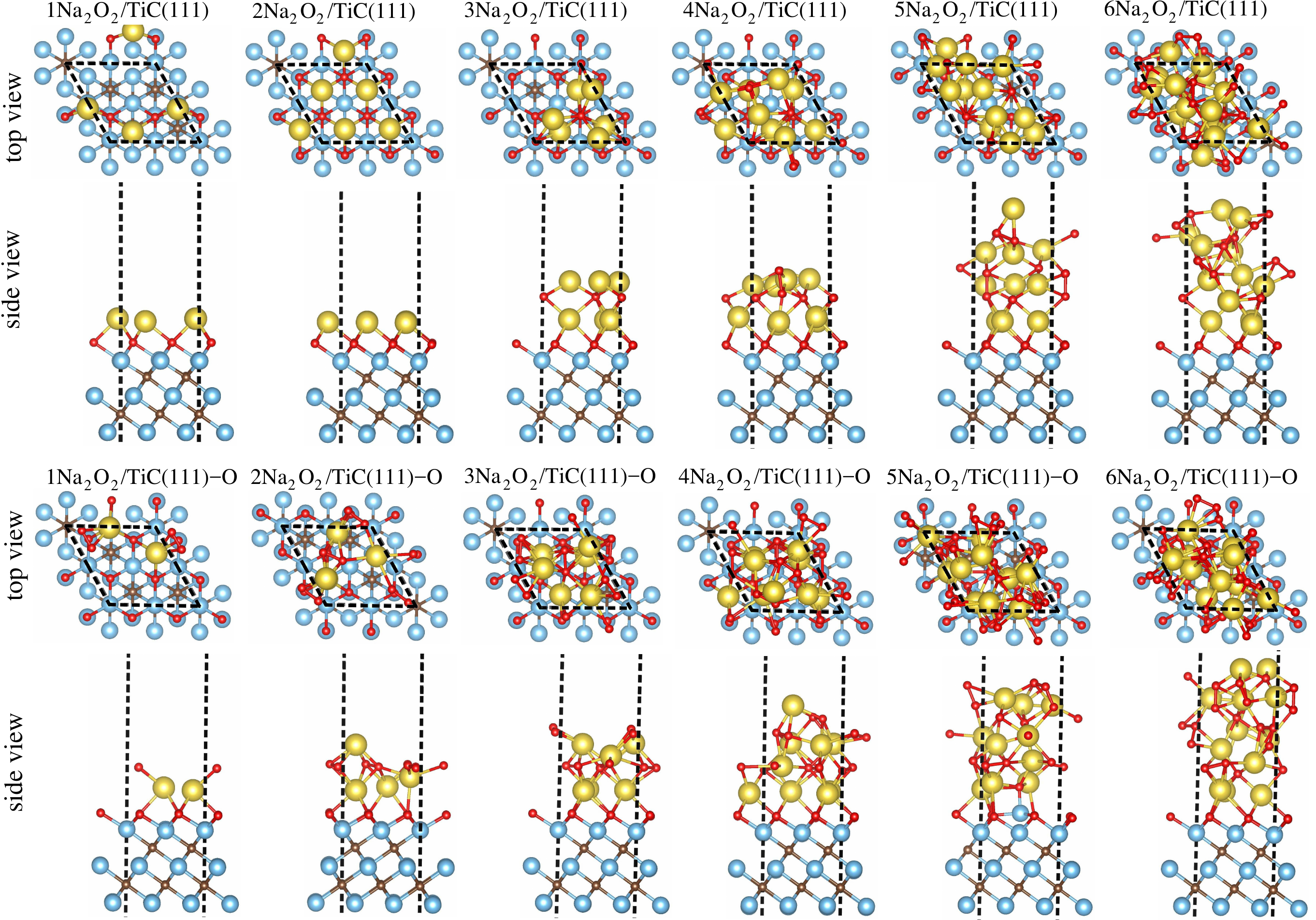}
\caption{\label{fig:Na2O2}Lowest energy configurations of $n$\ce{Na2O2} ($n$ = 1--4) molecules on
the non-oxidized and oxidized TiC(111). Blue, brown, red and yellow balls correspond to Ti, C, O and Na 
atoms, respectively. Ball sizes are drawn according to atomic radii.
}
\end{figure*}

\begin{figure*}
\includegraphics[width=0.9\textwidth]{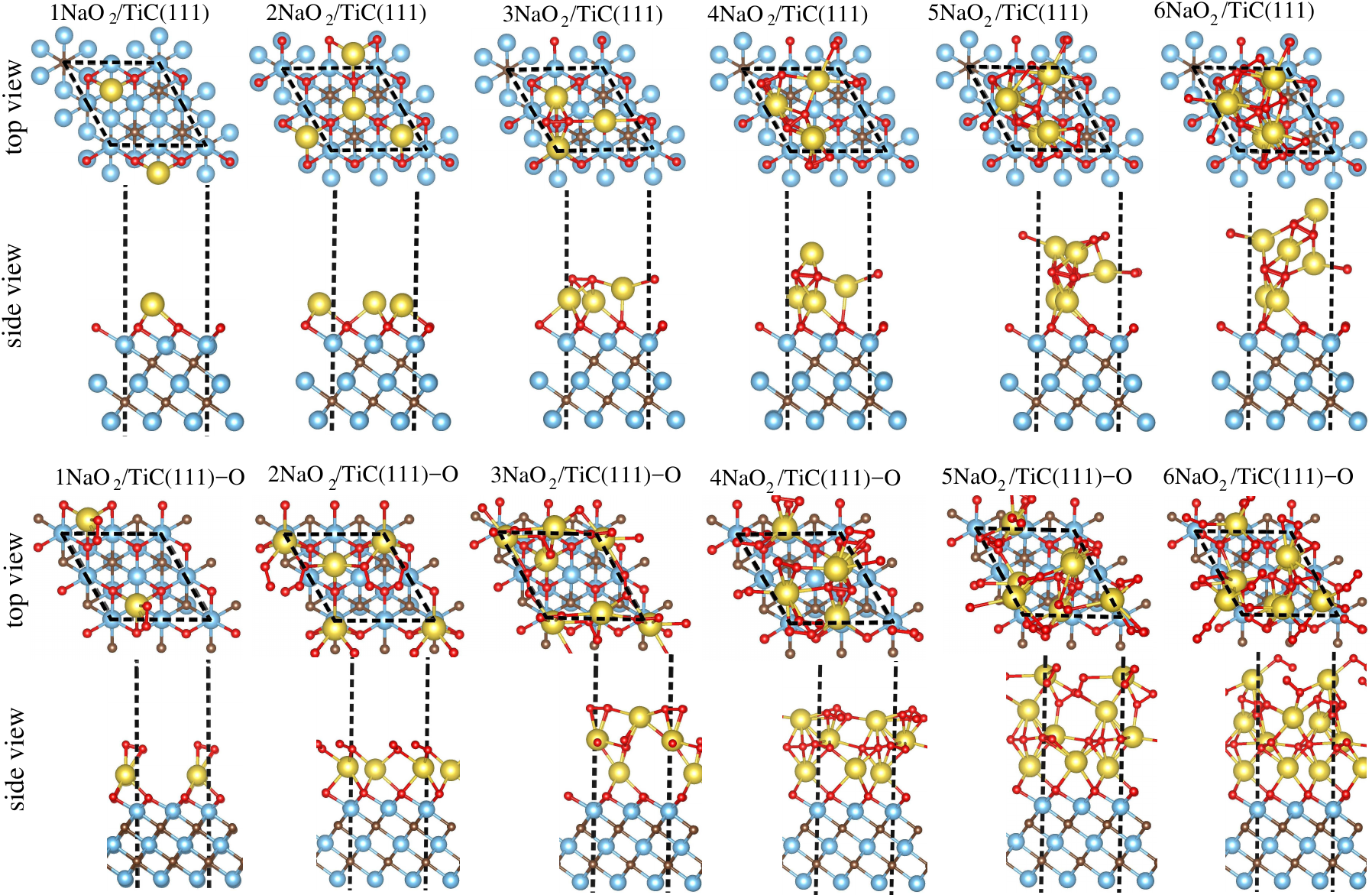}
\caption{\label{fig:NaO2}Lowest energy configurations of $n$\ce{NaO2} ($n$ = 1--4) molecules on
the non-oxidized and oxidized TiC(111), where blue, brown, red and yellow balls present Ti, C, O and Na
atoms, respectively. Ball sizes are drawn according to atomic radii.
}
\end{figure*}

We present the lowest energy structures of $n$\ce{M2O2} (M = Li, Na) and $n$\ce{MO2} (M = Na) molecules on
the pristine and oxidized TiC(111) surface in Figures \ref{fig:Li2O2}, \ref{fig:Na2O2} and \ref{fig:NaO2}.
In order to describe the structure of the molecules on the TiC(111) surfaces we measured the following
structural parameters: the molecular surface density, $\rho_{mol}$, calculated as the number of molecules per surface area,
the minimal Ti$-$O, M$-$O and M$-$M bond lengths, and the molecular layer thickness, $D$, calculated as
the vertical distance between the M atoms nearest and farthest from the surface, which might be helpful to
determine if the structure is crystal-like or distorted. All the structural parameters and
the adsorption energies of the lowest energy structures are given in Table \ref{tab:struc}, while the parameters
of both crystal-like and distorted structures can be found in Table 3 in the SI. The structural data for the corresponding
crystals and molecules is also provided in the SI for comparison.

The lowest energy structures of the adsorbed AMO molecules at $n$ = 1, 2 (with a molecular density of $\sim 0.03$ [1/\AA$^2$] and 0.06 [1/\AA$^2$] 
respectively) on the pristine TiC(111) surface form O and M atomic layers. In this structure the O atoms are at the hcp hollow positions of Ti atoms, taking the positions of the missing C atoms. The Li/Na atoms are on fcc hollow of C atoms, on the positions of the missing Ti atoms 
in the next atomic layer. The M$-$O bond lengths are close to the corresponding crystal bonds (see Table \ref{tab:struc}).
The formation of the layered structures due to the creation of O$-$Ti bonds and the complete break of the molecular O$-$O bonds
can be explained by the strong binding of the O atoms with the reactive Ti-terminated surface. Full molecular (2 molecules,0.06 [1/\AA$^2$])
coverage of the TiC(111) by \ce{2M2O2} and \ce{2MO2} leads to the expansion of the first O layer relative to a partial coverage, for example the Ti$-$O bond
lengths of \ce{2Li2O2}/TiC(111) increase by 0.18~{\AA} compared with \ce{1Li2O2}/TiC(111). Our finding is in a
good agreement with the structures obtained by {\it Wang et. al.} \cite{Wang2015} for \ce{1Li2O2} molecule on the 
TiC(111) surface.

At larger molecular density (0.09 [1/\AA$^2$] and 0.12 [1/\AA$^2$], which correspond to $n$ = 3 and 4) we found that molecules resemble a layered structure arrangement with M/O/M/O molecular layers on the TiC(111) surface,
in which the lowest O atoms stay at the hcp hollow positions on the TiC(111) surface, similarly to the case of $n$ = 1, 2,
however, the M and O atoms of the next molecular layers are slightly displaced from their ideal atomic hcp and fcc hollow positions.
The displacements are more significant for the $n$\ce{NaO2} molecules compared with the $n$\ce{Li2O2}
and the $n$\ce{Na2O2} molecules because of the different stoichiometry. 

An addition of the next two molecules ($n$ = 5 and 6 with the density of 0.15 [1/\AA$^2$] and 0.18 [1/\AA$^2$]) leads 
to large atomic displacements of the M/O/M layers, 
except for the first O layer bound to the TiC(111) surface. The M$-$O, O$-$O
and M$-$M bond lengths of the distorted structures are shorter compared with the ordered structures. 

To summarize this part, it is possible to say that initially the AMO atoms prefer to follow the atomic positions dictated by the TiC crystal. As more molecules are adsorbed, the structure changes in the direction of the native AMO crystal but more layers are needed to fully reach that. 

\begin{figure}
\includegraphics[width=0.45\textwidth]{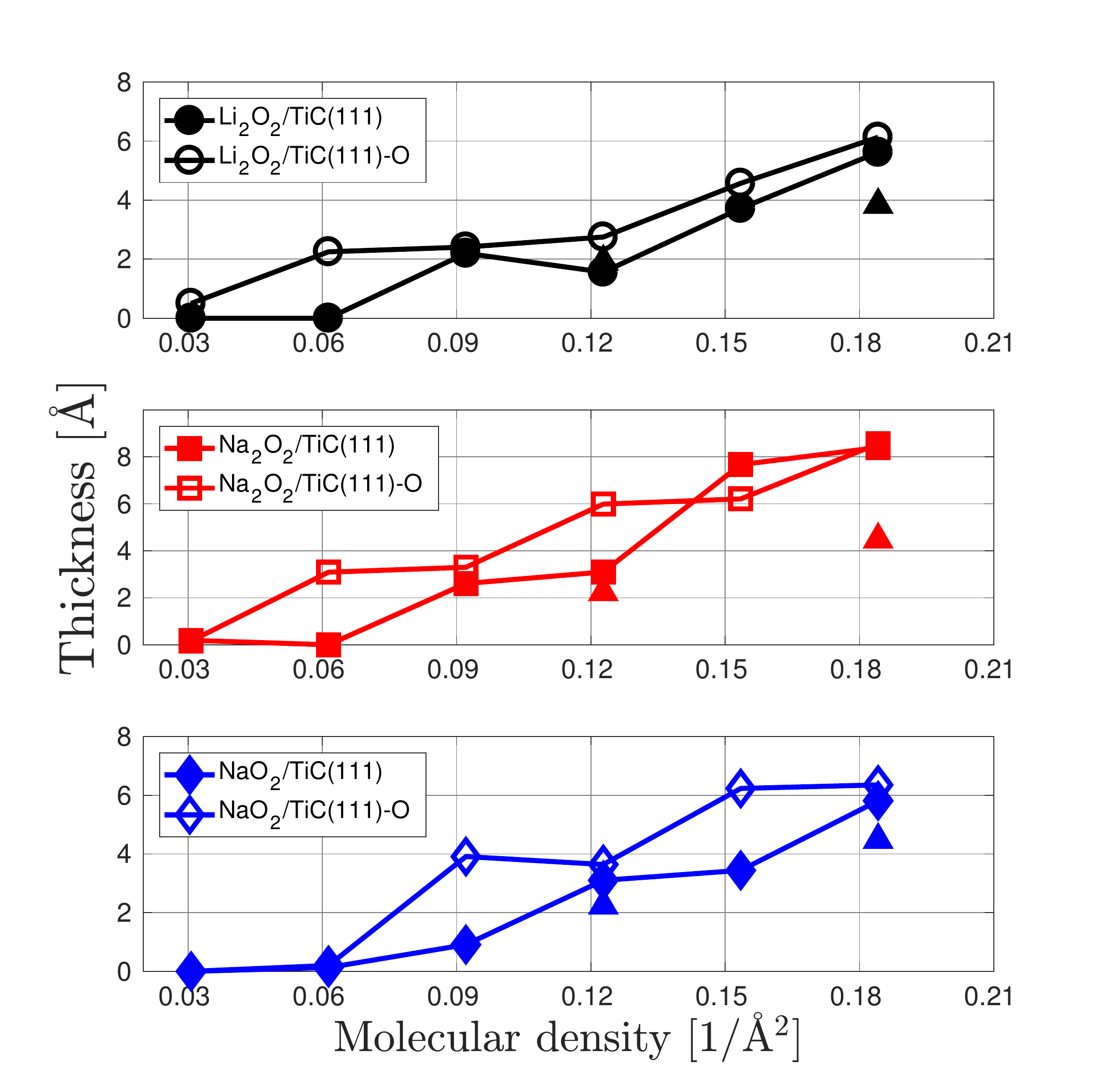}
\caption{\label{fig:thickness} The layers thickness for the lowest energy structures of $n$\ce{Li2O2},
$n$\ce{Na2O2} and $n$\ce{NaO2} molecules on TiC(111) and TiC(111)-O surfaces
vs molecular density. Triangles up and down correspond to the thickness of the crystals
in the original cell and compressed to the TiC(111) cell, respectively.
}
\end{figure}

The thickness, $D$, of the AMO molecular layers, is another parameter which can be compared with the respective $D$ of 
the corresponding compressed AMO crystal structure. Figure \ref{fig:thickness} shows the layers thickness of the 
lowest energy structures as a function of molecular density. We found that for $n$\ce{Li2O2}, $n$\ce{Na2O2},
and $n$\ce{NaO2} molecules on the pristine TiC(111) surface $D$ is near zero at $n$ = 1 and 2 showing that
in the first layer all metal atoms are at about the same height, while at $n$ = 3 and 4 $D$ in most cases change
only slightly with respect to the native crystal, namely by 12$-$19\%, except for \ce{4Na2O2} and \ce{3NaO2} systems which change by 38\% and 67\%, respectively. For the larger systems
(at $n$ = 5 and 6) the differences in $D$ between the adsorbed AMOs and the crystal structures are higher, for example 46\% for
\ce{6Li2O2}.

\begin{table*}
\caption{\label{tab:struc}
Adsorption energy, E$_{ad}$, and structural properties, such as minimal Ti$-$O, M$-$O and M$-$M bond lengths,
$d^{min}_{\text{Ti-O}}$, $d^{min}_{\text{M-O}}$ and $d^{min}_{\text{M-M}}$, respectively, of the $n$\ce{M2O2}
(M = Li, Na; $n$ = 1--6) and $n$\ce{NaO2} ($n$ = 1--6) molecules with molecular density, $\rho_{mol}$, and thickness,
$D$, on TiC(111) and TiC(111)-O surfaces. The data for crystal structure and molecules in gas-phase are also
provided for comparison. The adsorption energies are given in ~eV, while the bond lengths and thickness are
in ~{\AA}.
}
\begin{tabular}{cc|ccccc|ccccc}
$n$ & $\rho_{mol}$ &E$_{ad}$ & $d^{min}_{\text{Ti-O}}$ & $d^{min}_{\text{M-O}}$ & $d^{min}_{\text{M-M}}$ &
$D$ & E$_{ad}$ & $d^{min}_{\text{Ti-O}}$ & $d^{min}_{\text{M-O}}$ & $d^{min}_{\text{M-M}}$ & $D$ \\ \hline
  &     & \multicolumn{5}{c}{$n$Li$_2$O$_2$/TiC(111)} &  \multicolumn{5}{c}{$n$Li$_2$O$_2$/TiC(111)-O} \\ \hline
1 & 0.031 & $-10.00$  & 2.05 & 1.88 & 3.07 & 0.00 &  $-1.55$  & 2.02 & 1.88 & 2.98 & 0.50 \\
2 & 0.061 & $-9.74$  & 2.23 & 1.93 & 3.07 & 0.00 &  $-2.39$  & 2.03 & 1.89 & 2.27 & 2.25 \\
3 & 0.092 & $-8.71$  & 2.06 & 1.80 & 2.46 & 2.20 &  $-3.19$  & 2.02 & 1.86 & 2.43 & 2.41 \\
4 & 0.123 & $-7.86$  & 2.03 & 1.89 & 2.38 & 1.56 &  $-3.02$  & 2.05 & 1.77 & 2.78 & 2.75 \\
5 & 0.153 & $-6.50$  & 2.04 & 1.86 & 2.29 & 3.72 &  $-2.61$  & 2.02 & 1.82 & 2.39 & 4.57 \\
6 & 0.184 & $-5.87$  & 2.06 & 1.82 & 2.30 & 5.62 &  $-2.72$  & 2.06 & 1.86 & 2.22 & 6.13 \\
crys. &   &          &      & 1.98 & 2.65 & 1.92 &           &      & 1.98 & 2.65 & 1.92 \\
mol.  &   &          &      & 1.74 & 3.09 &      &           &      & 1.74 & 3.09 &      \\ \hline
  &     & \multicolumn{5}{c}{$n$Na$_2$O$_2$/TiC(111)}  & \multicolumn{5}{c}{$n$Na$_2$O$_2$/TiC(111)-O} \\ \hline
1 & 0.031 & $-10.80$ & 2.01 & 2.31 & 3.45 & 0.19 &  $-2.48$  & 2.01 & 2.30 & 3.34 & 0.16 \\
2 & 0.061 & $-10.08$ & 2.15 & 2.29 & 3.06 & 0.00 &  $-3.13$  & 2.05 & 2.22 & 3.09 & 3.09 \\
3 & 0.092 & $-8.88$  & 2.06 & 2.18 & 2.76 & 2.61 &  $-3.20$  & 2.06 & 2.22 & 2.99 & 3.30 \\
4 & 0.123 & $-7.40$  & 2.09 & 2.10 & 2.40 & 3.10 &  $-3.11$  & 2.06 & 2.19 & 2.85 & 5.99 \\
5 & 0.153 & $-6.50$  & 2.05 & 2.13 & 2.79 & 7.67 &  $-3.24$  & 1.91 & 2.16 & 2.76 & 6.21 \\
6 & 0.184 & $-5.89$  & 2.06 & 2.18 & 2.50 & 8.40 &  $-2.96$  & 2.05 & 2.13 & 2.55 & 8.56 \\
crys. &   &          &      & 2.31 & 3.05 & 2.24 &           &      & 2.31 & 3.05 & 2.24 \\
mol.  &   &          &      & 2.08 & 3.83 &      &           &      & 2.08 & 3.83 &      \\ \hline
  &     & \multicolumn{5}{c}{$n$NaO$_2$/TiC(111)} &  \multicolumn{5}{c}{$n$NaO$_2$/TiC(111)-O}  \\ \hline
1 & 0.031 & $-11.53$& 1.98 & 2.26 & 6.13  & $-$  &  $-1.45$  & 2.00 & 2.35 & 6.13 & $-$  \\
2 & 0.061 & $-10.81$& 2.03 & 2.25 & 3.54  & 0.13 &  $-2.15$  & 2.01 & 2.31 & 3.38 & 0.20 \\
3 & 0.092 & $-8.32$ & 2.05 & 2.23 & 3.24  & 0.91 &  $-1.92$  & 2.00 & 2.31 & 3.52 & 3.91 \\
4 & 0.123 & $-7.00$ & 2.05 & 2.21 & 3.11  & 3.10 &   $-2.18$ & 2.02 & 2.28 & 3.15 & 3.64 \\
5 & 0.153 & $-6.13$ & 2.02 & 2.25 & 3.06  & 3.44 &   $-2.17$ & 2.02 & 2.23 & 3.23 & 6.23 \\
6 & 0.184 & $-5.49$ & 2.03 & 2.23 & 3.14  & 5.81 &   $-2.21$ & 2.02 & 2.26 & 3.08 & 6.35 \\
crys. &   &         &      & 2.44 & 3.90  & 2.76 &           &      & 2.44 & 3.90 & 2.76 \\
mol.  &   &         &      & 2.13 & $-$   &      &           &      & 2.13 & $-$  &      \\ \hline
\end{tabular}
\end{table*}

\subsection{The effect of surface oxidation on the geometry of the adsorbed AMO layers}

The presence of an oxide layer on the TiC(111) surface, (TiC(111)-O), strongly affects the molecular
structures which differ significantly from that of the pristine surface. Despite the similar trends, such as breaking molecular O$-$O bonds and structural
reassembling, less ordered structures are already formed on the surface at $n$ = 1, as a result of no direct binding with the reactive  Ti atoms on the surface.

The layer thickness parameter tends to increase compared with the value of $D$ on the pristine TiC(111)
surface for all systems as a result of higher structural distortions. This trend can be observed already at $n$ =
1 and 2. One example is $D$ of 0.00~{\AA} and 2.25~{\AA} for \ce{2Li2O2} on the pristine and oxidized TiC(111) surfaces, respectively.
For higher molecular density the differences in $D$ still exist but show an oscillating pattern.

\subsection{d-band shift}

The analysis of changes in electronic properties can reveal additional details on the effects of molecular adsorption 
on clean and oxygen passivated surfaces. 

\begin{figure}
\includegraphics[width=0.45\textwidth]{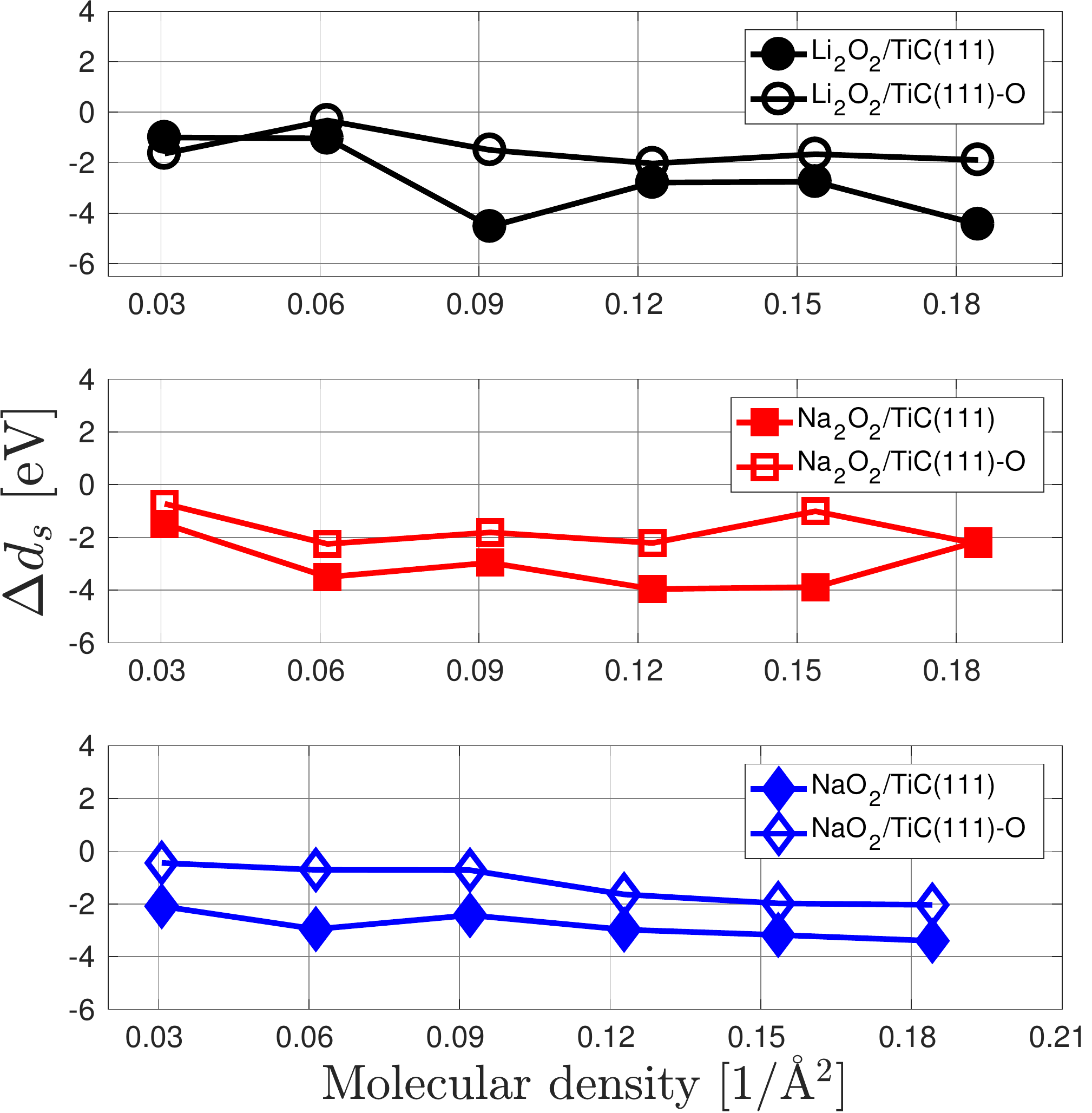}
\caption{\label{fig:d-band} The shift of the d-band center for \ce{Li2O2}, \ce{Na2O2} and
\ce{NaO2} on TiC(111) and TiC(111)-O surfaces vs molecular density per surface area.
}
\end{figure}

A shift of the center of the $d$ band ($d_s$) of the surface Ti atoms after adsorption can be considered
as an important parameter for the interaction of molecules with the Ti-terminated surface. On the basis of 
the $d$-band model, proposed by {\it Norskov et. al.} \cite{Hammer1995,Norskov2011}, $d_s$ can be used
in explaining the main trends in reactivity of the transition metals, which vary with element, surface 
structure and alloying \cite{Bligaard2007}. We present the calculated shifts, $d_s$, 
for the lowest energy systems in Figure~\ref{fig:d-band}. As can be expected, the shift in the center 
of the $d$ states on the TiC(111) surface is larger in most cases than the shift on the TiC(111)-O surface, which indicates 
stronger interaction of the molecules with the TiC(111) surface. For example, $\Delta d_s$ spreads from
$-1.49$~eV to $-3.96$~eV for the $n$\ce{Na2O2}/TiC(111) systems, while it ranges from $-0.72$~eV to $-2.25$~eV
for the $n$\ce{Na2O2}/TiC(111)-O systems.

Additional calculations of the local density of states (LDOS) for the lowest energy configurations are presented in the SI.

\subsection{Work function change}
\begin{figure}
\includegraphics[width=0.45\textwidth]{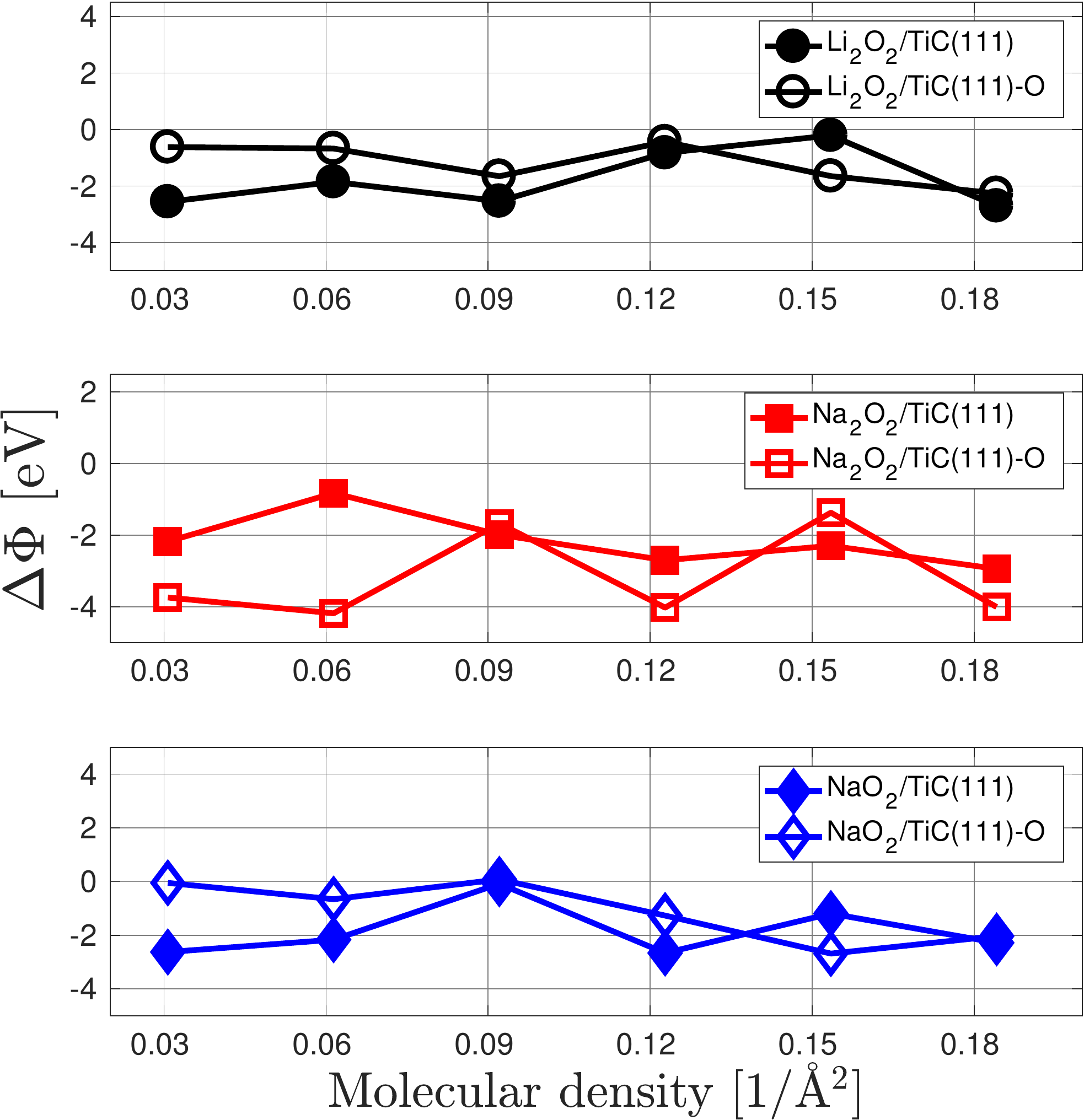}
\caption{The change in the work function for \ce{Li2O2},
\ce{Na2O2} and \ce{NaO2} on TiC(111) and TiC(111)-O surfaces.
}
\label{fig:Work-function-change} 
\end{figure}

Another surface property which can be calculated and measured is the change in the system work function as a function of molecular coverage. We
calculated the work function ($\Phi$) of the pristine and oxidized TiC(111) surfaces before and after adsorption,
and also the change in the work function of the systems,
namely - $\Delta \Phi = \Phi^{n\ce{M2O2}/TiC(111)} - \Phi^{TiC(111)}$. The work function changes are presented in Figure~\ref{fig:Work-function-change}. We also show the electrostatic
potentials of these systems, before and after adsorption, in the SI.

The value $\Phi$ of the pristine TiC(111) surface, calculated by us, is 4.57~eV, which is close to experimental
value of 4.7~eV \cite{Oshima1981}. An oxygen layer on the TiC(111) results in an increase of the work 
function to 5.20~eV. In most cases, the adsorption of the different AMO molecules leads to a reduction in the work
function - $\Delta \Phi$ is negative. This can be caused either by charge transfer or polarization of the molecules
on the surface or both. Examination of the changes in the work function shows that in all systems, there is no clear trend with coverage. This is to be expected as none of the grown crystals is polar and so it is logical to assume
that the work function will just oscillate as more layers are grown.

\section{Summary and Outlook}
In this work we have calculated the adsorption energies of varying coverage of \ce{Li2O2}, \ce{Na2O2}
and \ce{NaO2} molecules on clean and oxygen passivated TiC(111) surfaces. We showed that all the different
AMO molecules exhibit a similar behavior and that, as would be expected, adsorption on a clean surface
is initially much more favorable energetically than on an oxygen passivated surface. Furthermore, we showed that after
the deposition of two molecular layers, the adsorption energies at the clean and oxygen passivated surface approach one another
(see Fig. \ref{fig:adsorption_energy2}) and in fact the effect of the surface preparation becomes almost
unimportant. This is mainly because the newly adsorbed molecules are now lying on the previous AMO molecular
layers and not on the TiC itself, clean or oxygen passivated. To verify this we have compared the adsorption energies
to the crystal growth energy per layer for the respective crystals of \ce{Li2O2},\ce{Na2O2} and \ce{NaO2}.
We see that these energies are already very close. It should
be noted that for the oxygen passivated surface, which is more realistic experimentally, the adsorption energies are
close to that of the crystal growth energies already from the first molecule for all the AMO species that we checked.

The molecules we have checked have the option to adsorb at the surface or to self-assemble in solution and
form a cluster. Cluster formation energies are not equal to those of crystal growth and can larger at the beginning and
smaller at later stages where there are different facets for the formed clusters\cite{Kang2014}. As discussed before \cite{Johnson2014} the way  \ce{Li2O2} is crystallized, surface adsorption vs. solution based self-assembly, is strongly
affected by the solvent properties. However, the absolute surface adsorption energy may have some effect as well.
It is therefore interesting to map and compare those energies for different electrode surfaces and check their
possible relationship to cell performance.

\begin{acknowledgments}
This work was supported by the Planing \& Budgeting Committee of the Council  of High Education and the Prime Minister Office of Israel,  in the framework of the INREP project.
\end{acknowledgments}

\providecommand{\noopsort}[1]{}\providecommand{\singleletter}[1]{#1}%

\end{document}